\documentclass[aip,apl,reprint]{revtex4-1}

\usepackage{graphicx}

\DeclareTextSymbol{\degre}{OT1}{23}

\begin{document}

\title{Micro-Brillouin spectroscopy mapping of the residual density field induced by Vickers indentation in a soda-lime silicate glass}

\author{H.~Tran}
\author{S.~Cl\'ement}
\author{R.~Vialla}
\affiliation{Universit\'e Montpellier 2, Laboratoire Charles Coulomb UMR 5221, F-34095,Montpellier, France}
\affiliation{CNRS, Laboratoire Charles Coulomb UMR 5221, F-34095, Montpellier, France}

\author{D.~Vandembroucq}
\affiliation{Laboratoire de Physique et M\'ecanique des Milieux H\'et\'erog\`ene, PMMH UMR 7636 CNRS/ESPCI/Paris 6/Paris 7, 10 rue Vauquelin, F-75231 Paris cedex 05, France}
\affiliation{Laboratoire Surface du Verre et Interfaces, Unit\'e Mixte de Recherche CNRS/Saint-Gobain, 39 Quai Lucien Lefranc, F-93303 Aubervillers, France}

\author{B.~Ruffl\'e}\email{Benoit.Ruffle@univ-montp2.fr}
\affiliation{Universit\'e Montpellier 2, Laboratoire Charles Coulomb UMR 5221, F-34095,Montpellier, France}
\affiliation{CNRS, Laboratoire Charles Coulomb UMR 5221, F-34095, Montpellier, France}

\today
	
\begin{abstract}
High-resolution Brillouin scattering is used to achieve 3-dimensional maps of the longitudinal acoustic mode frequency shift in soda-lime silicate glasses subject to Vickers indentations. Assuming that residual stress-induced effects are simply proportional to density changes, residual densification fields are obtained. The density gradient is nearly isotropic, confirming earlier optical observations made on a similar glass. The results show that Brillouin micro-spectroscopy opens the way to a fully quantitative comparison of experimental data with predictions of mechanical models for the identification of a constitutive law.
\end{abstract}
\maketitle

Although mainly brittle, glasses can experience plastic deformation under high enough pressure or stress concentration ({\em e.g.} in the vicinity of a crack tip or below a sharp indenter). Not only plastic flow but also significant permanent densification are observed at small scale (from a few percents for soda-lime silicates to about twenty percents for amorphous silica). These phenomena and their important consequences for the mechanical strength have early been recognized~\cite{TAYL1949,MARS1964,ERNS1968,PETE1970}, but for reasons which are both theoretical and practical, the plastic behavior of glasses has long remained ill-understood. Indeed, the absence of a crystalline lattice does not allow using dislocations to explain plasticity. From the experimental side, the micrometer scale of the plastic zone makes difficult the measurements of deformations.

Recent theoretical, numerical, and technical developments, however, led to significant advances. From the modeling side, the description of plastic deformation of glassy materials as resulting from a series of local rearrangements of the amorphous structure~\cite{ARGO1979,FALK1998} has induced an intense numerical effort to simulate plastic flow at different time and length scales, see {\em e.g.}~\citet{RODN2011} for a recent review).

From the experimental side, the development in the field of instrumented nano-indentation~\cite{OLIV1992} has allowed to quantitatively probe the micro-ductile behavior of oxide glasses~\cite{SHOR1998}. The data gathered during such experiments are unfortunately far too limited to unambiguously characterize the mechanical response of the glass. Indeed, nano-indentation load-penetration curves reflect only the integrated response over the entire indented area. It is thus highly desirable to develop innovative micro-mechanics experiments which aim to image for example the local strain state of the glass surface after an indentation test. The comparison between such experimental data and finite element calculations is then a much more severe criterion for the constitutive law validation.

It was recently shown that spatially resolved optical spectroscopies can provide such information. Two-dimensional (2D) maps of the residual indentation-induced densification on a silica-sample surface were first obtained by Raman micro-spectroscopy using the D$_2$ line shift of the Raman spectrum of silica as a density marker~\cite{PERR2006}. This Raman band originates from the breathing of the three-membered rings~\cite{GALE1982} specific to silica. The rich set of data provided by the densification maps was then used to validate a new constitutive law describing the plastic deformation of silica at the micron scale~\cite{KERM2008}. More recently, \citet{PERR2011} characterized the residual densified area of a Vickers indent on a soda-lime silicate glass surface, using Cr$^{3+}$ luminescence micro-spectroscopy. The 2D density map revealed nearly circular iso-density contours, in sharp contrast with the star-shaped iso-density lines observed on silica~\cite{PERR2006}. Finite element calculations using the elastic-plastic constitutive equations developed for silica showed that the overall density gradient was well reproduced. On the other hand, the model clearly failed to describe the convex iso-density patterns found in the case of soda-lime silicate glass~\cite{PERR2011}. 

To identify a more general constitutive law, it seems useful to investigate other spectroscopies which could explore a wider range of glass compositions. To that effect, we propose to use high resolution Brillouin micro-spectroscopy. 3-dimensional (3D) maps of the Brillouin frequency shifts in the area beneath a Vickers micro-indent were obtained in a soda-lime silicate glass, demonstrating the potential of the technique.

Brillouin light scattering (BLS) is defined as inelastic scattering of light in a medium by thermally excited acoustical phonons. BLS is widely used in material science for measuring the bulk elastic properties (acoustic velocities) of small transparent samples. For pure backward Brillouin scattering in an isotropic medium, the measured Brillouin frequency shift $\delta\nu_{\rm B}$ is related to the refractive index $n$, the longitudinal sound velocity $v_{\rm L}$, and the incident vacuum wavelength $\lambda_0$, $\delta\nu_{\rm B} = 2 n v_{\rm L}/\lambda_0$. $v_{\rm L}$ is determined by the elastic properties and the density $\rho$ of the medium, $v_{\rm L}=\sqrt{M/\rho}$ where $M$ is the longitudinal elastic modulus. Generally, all these parameters are not independent and a quantitative determination of the moduli by BLS requires the knowledge of both $n$ and $\rho$.

High-resolution Brillouin spectra were recorded with a tandem spectrometer described elsewhere~\cite{VACH2006}. It consists of a four-pass planar Fabry-Perot (FP) interferometer in tandem with a spherical one, both controlled by laser light modulated at high frequency~\cite{SUSS1979}. The plane FP operates as a bandpass filter centered on the Brillouin line, reducing considerably the spurious elastic signal. The resolving unit is a scanned spherical FP of 25 mm spacing, {\em i.e.} with a free spectral range of about 3 GHz. This high-contrast/high resolution spectrometer is equipped with a microscope that permits experiments on small-size samples and Brillouin micro-cartographies with an improved spatial resolution.

\begin{figure}
\includegraphics[width=8.5cm]{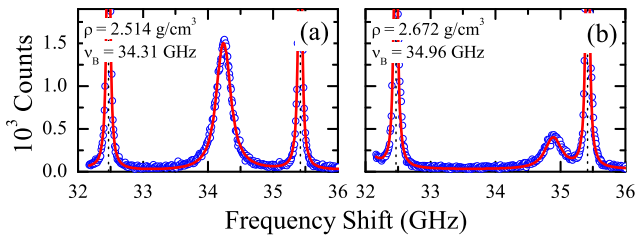}
\caption{(Color online) Brillouin spectra of the longitudinal acoustic modes at room temperature in the backscattering geometry for a pristine soda-lime silicate glass (a), and after a room-temperature hydrostatic compression at 20 GPa during 1 h (b). The dashed lines indicate the elastic peaks periodically transmitted.}
\label{Fig1}
\end{figure}

Light from a single mode Ar$^+$ laser operating at 514.5 nm is focused with a $\times$100 Plan Apo SL Mitutoyo objective of numerical aperture 0.55 on the sample surface. The latter is mounted on a computer-controlled XY piezoelectric stage. The scattered light is collected in the backscattering geometry. From these parameters, a lateral resolution of 1.2 $\mu$m is estimated. By monitoring the strength of the Brillouin line across the sample surface, an axial resolution of about 6 $\mu$m is obtained, in good agreement with the expected value. The laser power impinging on the sample is about 150 mW and a typical spectrum is recorded in 720 s. The sample was cut from a large plate of a standard window glass (Planilux, Saint Gobain Co., France), resulting in a parallelepiped of approximately 30$\times$30$\times$4 mm$^3$. Three series of five Vickers indents with loads of 0.5, 1 and 2 kg were prepared on the glass surface with an instrumented micro-indentation set-up.

\begin{figure}
\includegraphics[width=8.5cm]{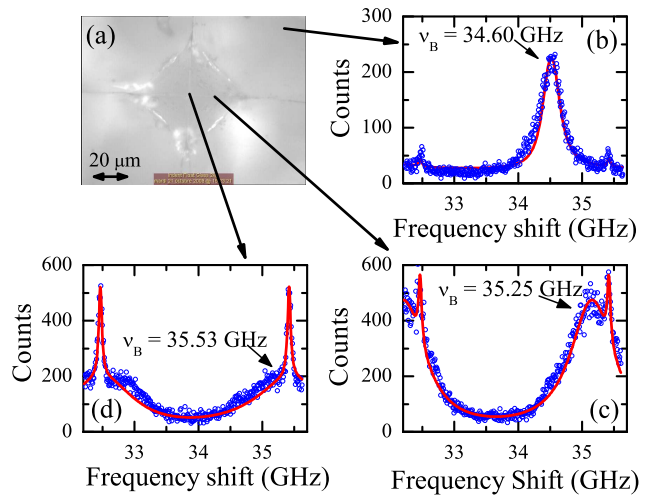}
\caption{(Color online) a) Top-view photograph of a 2-kg Vickers indent on a soda-lime silicate glass surface. b-d) Brillouin spectra taken at 5~$\mu$m beneath the surface, for a pristine sample volume far from the indented area, along the edge of the Vickers indenter at 12~$\mu$m from the center and at the center, respectively.}
\label{Fig2}
\end{figure}

As discussed above, the residual density field after indentation is expected to be a major signature of the plastic behavior of the glass. In view of a quantitative mechanical analysis of the BLS measurement, the calibration of the Brillouin frequency-shift variation with glass density is a crucial step. Not only the elastic modulus, but also the refractive index~\cite{DAND1998} exhibits a direct dependence on density. A possibility is to measure densified samples recovered from high pressure experiments which could serve as references. Densified samples of the same standard window glass were recently obtained using a multi-anvil high-pressure cell by \citet{JI2006}. After annealing for 2~h at the glass-transition temperature, small cylinders were maintained at high pressure and at room temperature for 1~h, and then slowly decompressed. For a maximum applied pressure of 20~GPa, the process resulted in a densification of 6.3\%, $\rho=2.672$~g.cm$^{-3}$, as compared to a pristine cylinder, $\rho=2.514$~g.cm$^{-3}$. The relative density change is indeed believed to saturate around 6\% for this glass composition~\cite{JI2006}. Both samples have been characterized by BLS to determine the elastic moduli~\cite{ROUX2010}. The high-resolution Brillouin spectra of the longitudinal acoustic modes in the backscattering geometry for the pristine and densified samples are shown in Fig.~\ref{Fig1}. The densification induces a well-defined increase of the Brillouin frequency shift, from 34.31~GHz for the pristine glass to 34.96~GHz for the densified one. Such a variation is well above the accuracy of the spectrometer which is better than 0.01~GHz. It clearly emphasizes the ability of BLS to monitor the densification in this glass composition.

Fig.~\ref{Fig2}a represents a top-view photograph of a 2-kg Vickers indented soda-lime silicate glass. Fig.~\ref{Fig2}b shows a Brillouin spectrum measured 5~$\mu$m below the surface at a sufficiently large distance from the indent to ensure that the pristine glass is probed. The Brillouin linewidth is comparable to the one observed in Fig.~\ref{Fig1}a, thus indicating a homogeneous scattering volume. However, the measured Brillouin frequency shift is 34.60~GHz at variance with the 34.31~GHz observed on the pristine sample of \citet{JI2006}. A slight difference in the commercial glass composition between lots might exist but it could hardly explain a variation of nearly 1\%. It is more likely that the Brillouin frequency shift change is simply due to the annealing treatment applied to the samples provided by \citet{JI2006}.

The Brillouin spectra obtained along the edge of the Vickers indenter at 12~$\mu$m from the center and at its center are shown in Figs.~\ref {Fig2}c and~\ref{Fig2}d, respectively. The experimental Brillouin linewidths are much broader in the stressed area. Indentation is indeed a highly inhomogeneous loading which modifies the glass beneath the surface to an inhomogeneous medium. Following the trend observed in Fig.~\ref{Fig1}, the measured Brillouin frequency shifts increase when the focal point is moved towards the center of the indent, where the maximum densification is expected. However, the magnitude of the measured variation, 0.93~GHz, is significantly larger than the 0.65~GHz increase associated to a 6.3\% densification derived from Fig.~\ref{Fig1}. This clearly shows that permanent densification can only account for part of the increase in the Brillouin frequency shift.

A natural explanation for that observation lies in the presence of residual (hydrostatic as well as deviatoric) stresses in the indented area.  Non-linear elastic effects associated with stresses indeed modify the local sound velocities probed by Brillouin spectroscopy via the third-order elastic constants~\cite{CAVA1999}. A fully quantitative analysis would imply computing the Brillouin frequency shifts induced by the indentation residual densification {\em and} stress fields and to compare it with the experimental BLS spatially resolved data. While such an analysis clearly goes beyond the scope of the present Letter, we stress that spatially resolved BLS spectroscopy appears as a very promising tool for a quantitative identification of a plastic constitutive law of oxide glasses.

It is, however, reasonable to assume that residual stress induced effects are simply proportional to densification. As a consequence, we consider in the following that a densification of 6.3\% is reflected by a linear increase of 0.93~GHz in the Brillouin frequency shift for that glass. Using this density gauge, the Brillouin frequency shifts measured throughout the indent can be converted to a local residual densification of the glass.

\begin{figure}
\includegraphics[width=8.5cm]{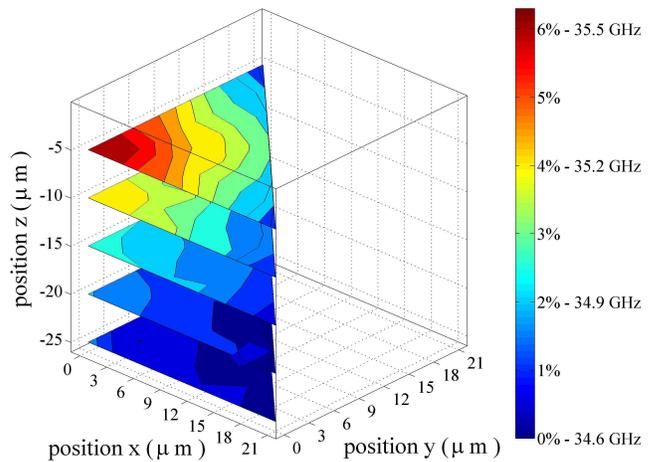}
\caption{(Color online) 3D mapping of the residual density field beneath a plastic impression left by a 2-kg Vickers indentation on a soda-lime silicate glass surface. The color scale indicates the residual densification in percent and the measured Brillouin frequency shifts, linearly related. $x$ and $y$ axes are along the diagonals of the indent and (0,0) defines its center. $z$ axis is normal to the pristine surface. The map represents a $x=y$ symmetrized quarter-view of the total indented area for visualization purpose but only a one-eight elementary unit of symmetry has been measured.}
\label{Fig3}
\end{figure}

2D maps, parallel to the pristine surface, were obtained by collecting 20 spectra covering one-eighth of the indent with a 3-$\mu$m spacing. 3D maps were made up by varying the focal point with a 5-$\mu$m spacing until the pristine glass was fully recovered. A representative 3D densification map of a 2-kg Vickers indent is shown in Fig.~\ref{Fig3}, using the Brillouin frequency shift of the longitudinal acoustic modes as a density gauge. The most densified area is found beneath the tip of the indenter and the plastic zone extends to about 25~$\mu$m from the glass surface. The density gradient is nearly isotropic, confirming earlier optical observations made on a similar glass~\cite{HAGA1984}. Iso-density contours at the surface show a spherical pattern, similar to those obtained using Cr$^{3+}$ luminescence micro-spectroscopy~\cite{PERR2011} and Raman micro-spectroscopy~\cite{DESC2011a}, hence validating the approach. We verified that the pattern was similar to those observed on other 2-kg Vickers indents. Densification maps were also obtained throughout 1-kg and 0.5-kg Vickers indents, revealing smaller circular densification gradients as expected.

In conclusion, BLS was used to obtain a first observation of the local densification induced by a Vickers test in the whole indented volume. It stresses the potential of Brillouin micro-spectroscopy to provide such detailed informations at the micron scale. As Brillouin spectroscopy is sensitive to residual stresses as well, it could be a promising tool to complement other spectroscopies, thus providing much richer data to be compared with constitutive model calculations. A second interest of BLS in the study of the elastic-plastic response of glasses lies is its applicability to a wide range of glass compositions. The density gauge does not rely on the existence of an optical mode specific to the glass structure or on the presence in the glass composition of transition or rare-earth ions. Hence, Brillouin micro-spectroscopy could be used to characterize the small scale mechanical response of amorphous solids in a continuous range of glass compositions exhibiting different plastic behaviors.

A clear limitation of the technique is the acquisition time needed to extract Brillouin frequency shifts. It severely hampers the spatial resolution of the densification map when keeping the total measurement time within reasonable limit. This issue should be eased by the recent development of a fast-acquisition Brillouin spectrograph based on a CCD and an original optical arrangement~\cite{VIAL2011}.

\begin{acknowledgments}
The authors would like to thank E. Barthel for enlightful discussions. This work was partially funded by the Agence Nationale pour la Recherche (Grant No. ANR-05-BLAN-0367-04, {\em PlastiGlass}) and R\'egion Languedoc-Roussillon (Omega Platform).
\end{acknowledgments}

%

\end{document}